\documentclass{PoS}
\usepackage{amsmath,amssymb,cite,epsfig,slashed,multicol,kotex}
\usepackage{tikz}
\usepackage[compat=1.1.0]{tikz-feynman}
\usetikzlibrary{shapes,arrows,positioning,automata,backgrounds,calc,er,patterns,trees,decorations.pathmorphing,decorations.markings}
\title{Dark matter direct detection with spin-2 mediators}
\ShortTitle{Dark matter direct detection with spin-2 mediators}
\author{Alba Carrillo-Monteverde\\ 
        Department of Physics and Astronomy, University of Sussex, Brighton BN1 9QH, UK\\        
        E-mail: \email{A.carrillo-Monteverde@sussex.ac.uk}}
\author{\speaker{Yoo-Jin Kang}\\
       Department of Physics, Chung-Ang University, Seoul 06974, Korea\\
       E-mail: \email{yoojinkang91@gmail.com}}
\author{Hyun Min Lee\\ 
        Department of Physics, Chung-Ang University, Seoul 06974, Korea\\
        E-mail: \email{hminlee@cau.ac.kr}}
\author{Myeonghun Park\\ 
        Center for Theoretical Physics of the Universe, Institute for Basic Science, Daejeon 34051, Korea\\
        Institute of Convergence Fundamental Studies and School of Liberal Arts, Seoul National University of Science and Technology, Seoul 01811, Korea\\
        E-mail: \email{ishaed@gmail.com}}
\author{Veronica Sanz\\ 
        Department of Physics and Astronomy, University of Sussex, Brighton BN1 9QH, UK\\
        E-mail: \email{v.sanz@sussex.ac.uk}}                        
\abstract{We consider models where a massive spin-two resonance acts as the mediator between Dark Matter (DM) and the SM particles through the energy-momentum tensor. We examine the effective theory for fermion, vector and scalar DM generated in these models and find novel types of DM-SM interaction never considered before. We identify the effective interactions between DM and the SM quarks when the mediator is integrated out, and match them to the gravitational form factors relevant for spin-independent DM-nucleon scattering. We also discuss the interplay between DM relic density conditions, direct detection bounds and collider searches for the spin-two mediator.}
\FullConference{ICHEP 2018, International conference on high energy physics\\
		COEX, Seoul, Republic of Korea.\\
		July, 4--11 2018}	
\begin{document}
\section{Introduction}
Dark Matter (DM) is known from many observations such as galaxy rotation curves, gravitational lensing effects, Cosmic Microwave Background (CMB) and so on. Planck satellite has observed the CMB and inferred that DM accounts for $\Omega h^2\simeq 0.1186$ \cite{Planck}.
Among the DM experiments, direct detections give strong bounds \cite{DDs}. 
The effective field theory approach is suitable to the scattering for DM and nucleons in direct detection.
Moreover, the same effective interactions at the freeze-out of DM annihilation in the early Universe and at the collider experiments are valid. Until now, the spin-0 and spin-1 mediators have been mainly proposed.
We consider the spin-2 mediated dark matter model. The interactions are written through the energy-momentum tensor \cite{hml14,hml13}. We identify the effective DM-nucleon scattering interactions by integrating out the mediator and match from quarks to nucleons with gravitational form factors. We examine the differential scattering event rate for DM-nucleon scattering with non-relativistic effective operators and discuss the relic density condition of DM and experimental bounds. 
\section{Spin-2 mediator and dark matter}
We introduce the interaction of a massive spin-2 particle (${\cal G}_{\mu\nu}$) to the SM particle and DM with the energy-momentum tensor, \cite{hml14,hml13}
\begin{equation}
{\cal L}_{\rm int}=-\frac{c_{\rm SM}}{\Lambda}{\cal G}^{\mu\nu}T^{\rm SM}_{\mu\nu}-\frac{c_{\rm DM}}{\Lambda}{\cal G}^{\mu\nu}T^{\rm DM}_{\mu\nu}.
\end{equation}
In this model, we consider the tree-level scattering amplitude between DM and SM quarks through a massive spin-2 propagator ${\cal P}_{\mu\nu,\alpha\beta}=( G_{\mu\alpha}G_{\nu\beta}+G_{\nu\alpha}G_{\mu\beta}-\frac{2}{3}G_{\mu\nu}G_{\alpha\beta} )/2$ with $G_{\mu\nu}=\eta_{\mu\nu}-(q_\mu q_\nu)/m_G^2$.
The tensor ${\cal P}_{\mu\nu,\alpha\beta}$ satisfies the traceless and transverse condition for on-shell mediator \cite{hml14}.
The massive spin-2 particle is integrated out and that leads to the effective amplitude described as traceless part and trace part of the energy-momentum tensors, with $\tilde{T}_{\mu\nu}=T_{\mu\nu}-\frac{1}{4}\eta_{\mu\nu}T$,
\begin{equation}
{\cal M}=\frac{i c_{\rm DM}c_{\rm SM}}{2m_G^2 \Lambda^2}\big( 2\tilde{T}^{\rm DM}_{\mu\nu}\tilde{T}^{{\rm SM},\mu\nu} -\frac{1}{6}T^{\rm DM}T^{\rm SM}  \big).
\end{equation}
 The energy momentum tensors for the SM particles and DM are introduced in the \cite{hml14,hml13, Orig}. In this paper, we consider only quark$(\psi)$ coupling for SM part.
In the case of scalar operator, it takes scalar form factor while twist-2 operator takes gravitational form factor
\begin{equation}
\langle N(p_2)| T^\psi |N(p_1)\rangle =-F_S(q^2)m_N\bar{u}_N (p_2)u_N (p_1)  
\label{sform}
\end{equation}
\begin{equation}
\langle N(p_2)| \tilde{T}^\psi_{\mu\nu} |N(p_1)\rangle = F_T(q^2)\tilde{T}^N_{\mu\nu}
\label{gform}
\end{equation}
with $q=p_1-p_2$. In the case of the twist-2 operator of quarks, there can be more gravitational form factors but others can be zero in a holographic description of QCD in a five-dimensional AdS spacetime \cite{Calson}. If we assume that the momentum transfer is zero, the scalar form factor becomes just mass fraction of light quark in a nucleon, $F_S(0)= f_{T\psi}^N$,  and gravitational form factor becomes the second moments of PDF, $F_T(0)=\psi(2)+\bar{\psi}(2)=\int_0^1 dx\ x(\psi(x)+\bar{\psi}(x))$ \cite{Drees,ishi}.
\section{Model constraints and results}
The effective operators consist of the complete set of 4 Hermitian quantities, momentum transfer$(\vec{q})$, relative perpendicular velocity$(\vec{v}^\perp\equiv \vec{v}+\vec{q}/\mu_N )$, spin of DM$(S_\chi)$ and spin of nucleon$(S_N)$ where $\mu$ is reduced mass of DM and nucleon. The non-relativistic effective operators are given in \cite{Fitz13}.
In the case of fermion DM$(\chi)$, we note that the effective amplitude has the dimension-8 operator due to the massive spin-2 mediator \cite{Orig}. Also, we figure out the leading order of effective Lagrangian for DM and nucleon scattering is independent of dark matter spin, where ${\cal O}_1^{\rm NR}=1$,
\begin{equation}
{\cal L}_{\rm eff}\simeq \frac{c_{\rm DM}c_\psi m_{\rm DM}^2m_N^2}{2m_G^2 \Lambda^2}\bigg(6F_T-\frac{2}{3}F_S \bigg){\cal O}_1^{\rm NR}.
\end{equation}
\begin{figure}[h!]
  \centering
\includegraphics[width=0.45\textwidth]{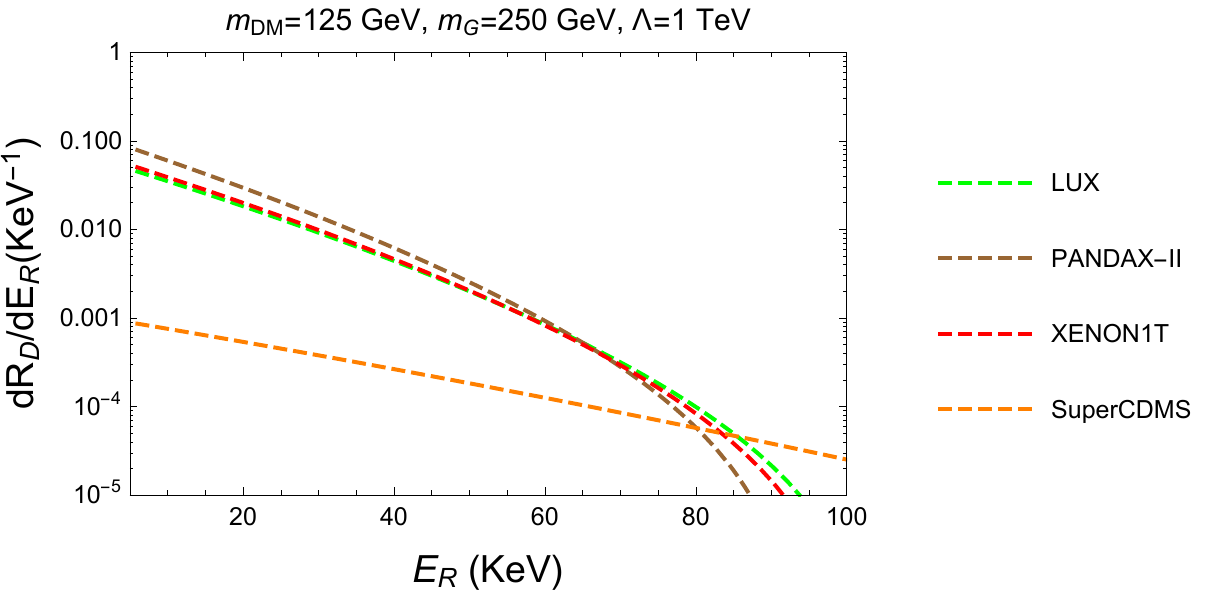}
\includegraphics[width=0.45\textwidth]{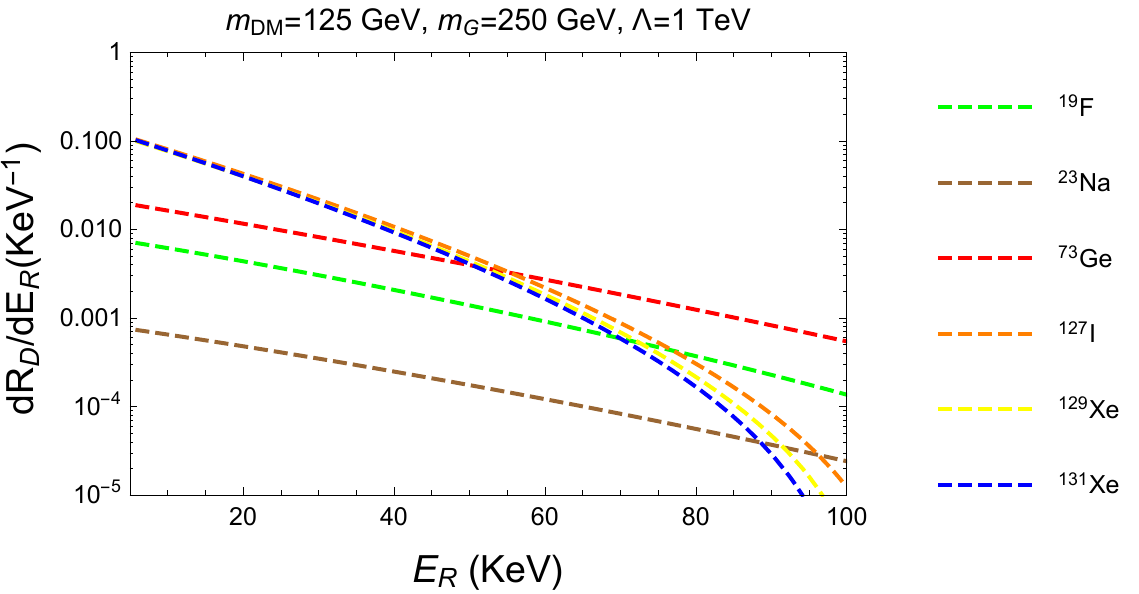} 
\caption{The differential event rates in current(left) and mock(right) experiments for fermion DM}
\label{differential}      
\end{figure}
We compute the differential scattering event rates using the mathematica package called DMFormFactor \cite{Fitz13,Fitz15} taking benchmark points satisfying relic density and constraints. We use the parameters for various current experiments and mock experiments for fermion DM case in Fig. \ref{differential}.
\begin{figure}
\centering
\includegraphics[width=0.32\textwidth]{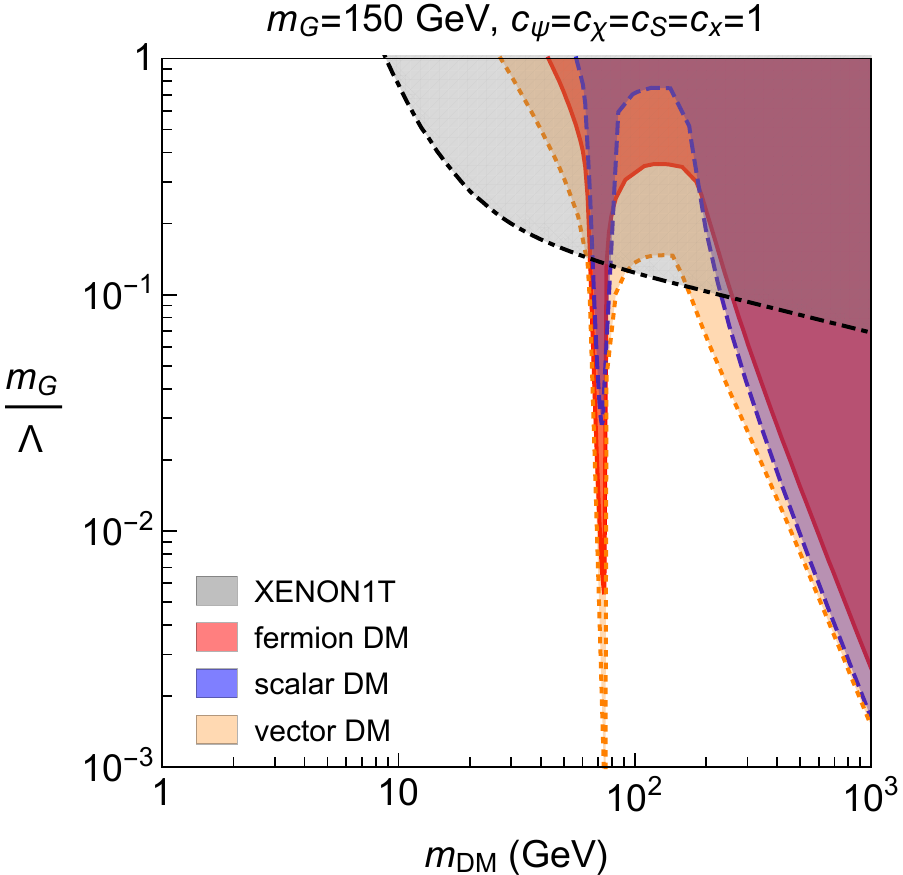} \ 
\includegraphics[width=0.32\textwidth]{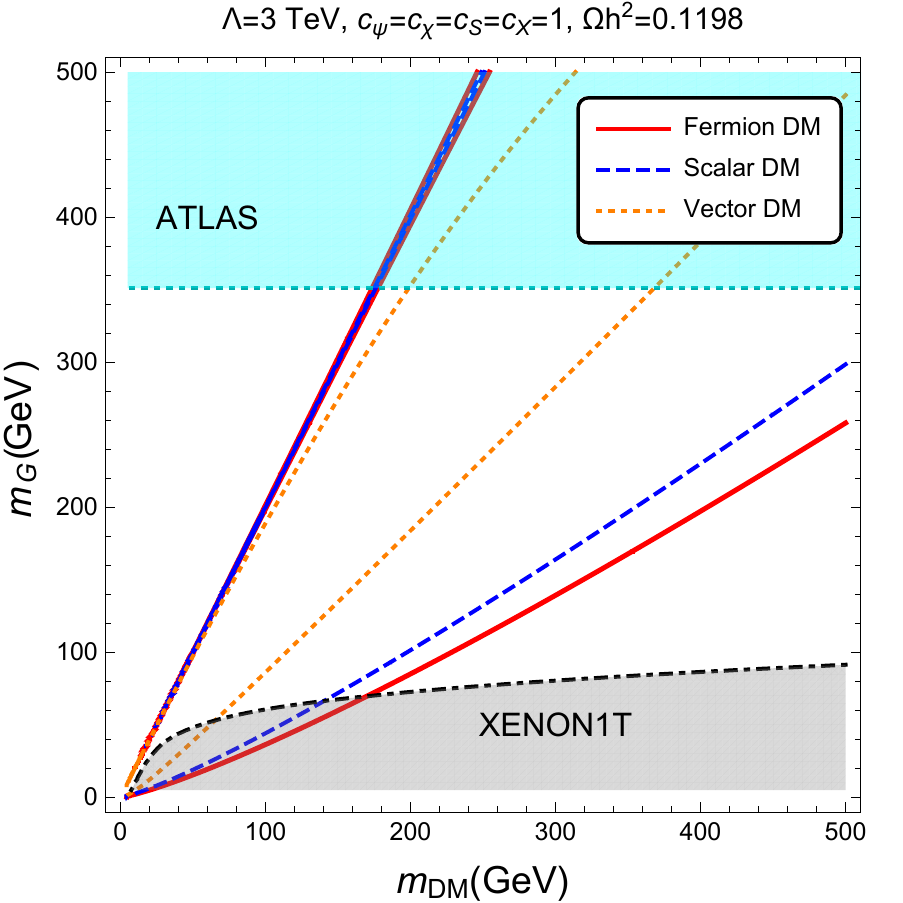} 
\caption{Parameter space with constraints where $m_{\rm DM}=150$ GeV and $c_{\psi}=c_{\rm DM}=1$.}
\label{relic}      
\end{figure}
Dark matter relic density is determined by annihilation cross sections for DM, DM$\rightarrow \psi\bar{\psi}$ or $GG$ and the formulae are in \cite{hml14,hml13, Orig}. 
In figure \ref{relic}, we show the parameter space for $m_{\rm DM}$ vs $m_G/\Lambda$ with $m_G=150$ GeV (left) and $m_{\rm DM}$ vs $m_G$ with $\Lambda=3$ TeV where $c_\psi=c_{\rm DM}=1$ in both sides. The DM relic density over closes the universe in red, blue and orange for fermion, scalar and vector DM, respectively. We consider XENON1T \cite{XENON} with gray and ATLAS dijet bounds \cite{ATLAS} with cyan. In particular, taking a zero momentum transfer for the DM-nucleon scattering, we obtain the spin-independent scattering cross section with the results of eq. (\ref{sform}) and (\ref{gform}) \cite{Orig, Rizzo}, 
\begin{equation}
\sigma^{\rm SI}_{{\rm DM}-A}=\frac{\mu_A^2}{\pi}\big( Zf_{p}^{\rm DM}+(A-Z)f_n^{\rm DM} \big)^2
\end{equation}
with
\begin{equation}
f_{n,p}^{\rm DM}=\frac{c_{\rm DM}m_N m_{\rm DM}}{4m_G^2 \Lambda^2}\Big(\sum_{\psi=u,d,s,c,b}3c_\psi (\psi(2)+\bar{\psi(2)})+\sum_{\psi=u,d,s}\frac{1}{3}c_\psi f_{T\psi}^{n,p} \Big).
\end{equation}
We also considered the light DM ($m_{\rm DM} \lesssim 10$ GeV) with light DM experiments in the paper \cite{Orig} and found that it is strongly constrained by DarkSide-50 experiment \cite{DS50}.
\section{Conclusion}
We have presented the effective interactions between DM and the SM quarks with spin-2 mediator. We have shown the differential event rates for DM-nucleon scattering and imposed the bounds from direct detection, relic density condition as well as LHC dijet searches on to the parameter space. In this work, we consider only quark coupling but gluon coupling contribution will be shown in the future work.
\section{Acknowledgment}
The work of YJK and HML is supported in part by Basic Science Research
Program through the National Research Foundation of Korea (NRF) funded by the Ministry of Education, Science and Technology (NRF-2016R1A2B4008759). The work of YJK was supported by IBS under the project code, IBS-R018-D1. ACM is supported by the
Mexican National Council for Science and Technology (CONACyT) scholarship scheme.
\end{document}